\documentclass[reqno]{amsart}
\usepackage{graphicx,amsmath}
\usepackage{epstopdf}
\usepackage{float}
\usepackage[utf8]{inputenc}
\usepackage{textcomp}

\def\bq{\mathbf q}
\def\bQ{\mathbf Q}
\def\bJ{\mathbf J}
\def\bb{\mathbf b}
\def\bB{\mathbf B}
\def\re#1{(\ref{#1})}
\def\f{\frac}

\begin{document}

\title{Models of ballistic propagation of heat  at low temperatures}

\author{R. Kov\'acs$^{123}$ and P. V\'an$^{123}$}

\address{
$^1$Department of Theoretical Physics, Wigner Research Centre for Physics,
Institute for Particle and Nuclear Physics, Budapest, Hungary and 
$^2$Department of Energy Engineering, BME, Budapest, Hungary and 
$^3$Montavid Thermodynamic Research Group, Budapest, Hungary}

\date{\today}

\begin{abstract}
Heat conduction at low temperatures shows several effects that cannot be described by the Fourier law. In this paper, the performance of various theories is compared in case of wave-like and  ballistic propagation of heat pulses in NaF.  
\end{abstract}
\maketitle
 
\section{Introduction}
The detection and modeling of non-Fourier effects can be vital in cases of energy management of very fast processes, in micro-electro-mechanical systems, and also in materials with heterogeneous substructures \cite{BotEta16a}. The second sound -- the wave-like propagation of heat -- is observed at low temperatures in various materials, such as NaF, Bi and superfluid He. The phenomenon was predicted by Tisza and Landau \cite{Tisz38a,Lan41a} and first measured by Peshkov \cite{Pes44a}. Its simplest theory is a memory extension of the Fourier equation, the so-called Maxwell--Cattaneo--Vernotte (MCV) equation \cite{Cat48a,Ver58a1,Max867a}. Later, the microscopic model of Guyer and Krumhansl  \cite{GuyKru66a1} proved extremely helpful in experimental observations of the second sound in solids, and plays an increasingly important role in recent studies \cite{SelEta16b,Zhu16a}. The ballistic-type propagation is more complicated, to observe, the presence of such effect is very sensitive to boundary conditions. Several experiments have detected the latter phenomenon \cite{JacWalMcN70, JacWal71}, and various theories have been developed to provide explanation and  modeling  \cite{Rog71a, Rog72a,DreStr93a,Chen01,Ma13a}. A short review of the interaction of experiments and theory of heat conduction beyond the Fourier law is given in \cite{Van15a}. 

In this paper, we analyze the low-temperature NaF experiments of Jackson, Walker and McNelly, and compare the performance of theories in reproducing the experimental data. Our primary concern and reference point is the identification of material parameters in our continuum theory, developed in \cite{KovVan15a}. This theoretical framework covers all the known model candidates, including the nine-field equations of Extended Thermodynamics, and is based only on universal principles. In \cite{KovVan15a}, we have used artificial parameters in order to demonstrate the phenomena,  while in this paper we provide a detailed analysis and modeling.

\section{NaF experiments}

\begin{figure}[h]
\centering
\includegraphics[width=13cm,height=13cm]{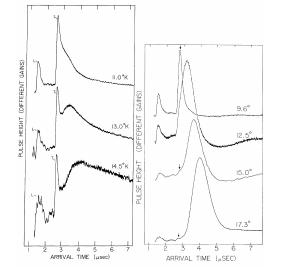}
\caption{Original results of NaF experiments: the right one was published in  \cite{JacWalMcN70} and the left one in \cite{JacWal71}. L and T denote the longitudinal and transversal peaks of ballistic propagation, respectively. }
\label{fig:bal1}
\end{figure}

In the experiments analysed here, a heat pulse is applied at the front end of the sample,  and the relevant measurement data is the rear-side temperature as the function of time. In figure \ref{fig:bal1}, one can see several data series for two different samples and initial temperatures \cite{JacWalMcN70,JacWal71}. Our focus will be on the middle curve on the left-hand side, which shows the characteristic behaviour the best.

In what follows, we summarize the available information about the parameters and  subsidiary conditions of the experiments, which are crucial for theoretical  reproduction.  

\subsection{Identification of the samples} The papers mention at least six different samples, with different sizes and purities. Five of them are mentioned in \cite{JacWal71} and one in \cite{McNEta70a}. The crystals can be identified only by their length and by impurity classes such as ``very pure'', ``less pure'', etc. This twofold identification is crucial when we are seeking for the proper parameters of the respective measurement data.

\subsection{The material parameters} {\it Thermal conductivity} depends on temperature and on the crystal impurity. Walker \cite{Wal63} presents the thermal conductivity curves for several crystals, the relevant ones being NaF \#1 and \#2. The peak value of the conductivity of \#1 is approximately the double of the peak value of the second one. This sample has not been used in further measurements, in spite of the seemingly better properties.

Thermal conductivity is not the only relevant material parameter, as one needs to use  {\it density} and {\em specific heat} as well. The temperature dependency of thermal conductivity is well-discussed in the literature. Bargmann and Steinmann \cite{BarSte05a} give the value of density and of specific heat according to the Gmelin Handbook \cite{Gmelin93}, but those values are different of any of the above-discussed NaF crystals. Hardy and Jaswal \cite{HarJas71a} have calculated the temperature dependency of specific heat, which changes significantly between 9 K and 17 K. This property is analyzed by Coleman and Newmann \cite{ColNew88}, too. However, one cannot find information regarding the crystal impurity dependency of the specific heat.

The {\it relaxation times} are determined from the measurement via fitting to the model equations, hence, different models can lead to different relaxation times. Ballistic speed is not an independent further parameter: it can be computed from the respective theory once the value of relaxation time is known.

\subsection{Boundary conditions} There are two important questions about the boundary conditions: How long is the input pulse? Can the rear end be adiabatic? 

According to McNelly et al. \cite{McNEta70a} the pulse length is between $0.1 \hbox{\textmu} s$ and $1 \hbox{\textmu} s$. The theoretical analyses use various values within this interval. Dreyer and Struchtrup apply the lower limit, $0.1 \mu s$ \cite{DreStr93a}. Yanbao Ma used two different values, $0.24\, \hbox{\textmu}s$ and $0.36 \, \hbox{\textmu}s$, for the evaluation of the same experiment \cite{Ma13a1}. 

Regarding the other question, Jackson et al. \cite{JacWal71} do not provide any direct information about the rear boundaries. The flattening of the measured curves in Figure \ref{fig:bal1} indicates that the rear side cools back non-negligibly during th eitme scale of the measurement. Theoretical modeling did not use this observation.

\subsection{Transversal or longitudinal?}

The presence of shear and longitudinal ballistic propagation indicates that the experiment cannot be modeled in one spatial dimension. Moreover, the sketch of McNelly regarding  experimental arrangement  in \cite{McN74t}  confirms that, as the specimen is cubelike and the heat excitation affects only the middle part of the front face. However, model calculations are mostly one dimensional \cite{DreStr93a,KovVan15a}, except for \cite{Ma13a1}. Moreover, usually the transversal, rather then the longitudinal,  ballistic peak is reproduced in these one dimensional models. More properly, something between the two is provided, corresponding to the propagation speed calculated with the one dimensional elasticity parameter, the Young modulus.

On the one hand, it is the longitudinal ballistic propagation that
can be connected to one-dimensional models, in the form of volumetric thermoelastic effects \cite{FriCim98m}, while the transversal one cannot be interpreted within such framework. On the other hand, the signal speed may be calibrated by the proper material parameters in both cases. Hence, in this work we choose the simultaneous reproduction of the peak of the transversal ballistic propagation, and of the with the second sound, in order to keep our results comparable to results of other theories.

\section{Non-equilibrium thermodynamics}

In the present study the NaF experiments are modeled in the continuum framework of  non-equilibrium thermodynamics. This is the starting point regarding the comparison with other models. Therefore, the derivation of our model is briefly discussed. Further details are given in \cite{KovVan15a}.

Let us consider a rigid conductor, where mass density is constant and material time derivatives are identical to partial time derivatives. We start with the balance of internal energy,
\begin{equation}
\partial_t e + \nabla\cdot \mathbf q =0,
\label{inten_bal}
\end{equation}
where $\partial_t$ is the partial time derivative, $e$ denotes the density of internal energy, $\nabla \cdot$ stands for the divergence, and $\mathbf q$ is the heat flux. The second law reads as
\begin{equation}
\partial_t s + \nabla\cdot \mathbf J \geq 0,
\label{s_bal}
\end{equation}
where $s$ denotes entropy density and $\mathbf J$ is the entropy current density. The assumption of local equilibrium of thermal interaction leads to Fourier's equation. Non-equilibrium thermodynamics can characterize deviation from local equilibrium both in the entropy density and in the entropy flux by introducing internal variables (dynamical degrees of freedom \cite{Verhas97})  and current multipliers \cite{Nyi91a1}, respectively. In case of heat conduction, it is customary to choose heat flux,  $\mathbf q$, as internal variable \cite{MulRug98b,SelEta16b}. We also introduce a second order symmetric tensor, $\mathbf Q$, as the next approximation. Thermodynamic stability is preserved by assuming a quadratic deviation in the entropy density \cite{Verhas97}:
\begin{equation}
s(e, \bq, \bQ) = s_{\text{eq}}(e) - \frac{m_1}{2} \bq\cdot \bq- \frac{m_2}{2} \bQ : 
\bQ,
\label{neqs}
\end{equation}
where  $m_1$ and $m_2$ are positive material parameters. This choice of field variables allows comparison with Extended Thermodynamics. The entropy current is considered as
\begin{equation}
\bJ = \bb\cdot\bq + \bB :\bQ,
\label{neqJ}
\end{equation}
where $\mathbf b$ is a $2^{\text{nd}}$ order tensor and $\mathbf B$ is a $3^{\text{rd}}$ order tensor. These quantities are called current multipliers -- they were first introduced by Nyíri \cite{Nyi91a1} -- and they are constitutive functions. Then we can calculate the entropy production:
\begin{eqnarray}
\partial_t s+ \nabla\cdot \bJ 
&=& \left(\bb - \frac{1}{T} \mathbf I \right): \nabla\bq + 
	\left(\nabla\cdot\bb -  m_1 \partial_t\bq\right)\cdot \bq + \nonumber\\
& &	\left(\nabla\cdot\bB -  m_2 \partial_t\bQ\right): \bQ +
	\bB \vdots \nabla\bQ\,\, \geq\;\, 0. 
\label{entrpr}
\end{eqnarray}
Here, $\mathbf I$ stands for the identity tensor and the triple dot denotes the full contraction between the third order tensors.  One can separate the thermodynamic forces and fluxes according to Table 1, and introduce  linear relationship between them. 
\begin{center}
\begin{tabular}{c|c|c|c|c}
       &Thermal & Extended thermal & Internal & Extended internal \\ \hline
Fluxes & $\nabla\cdot \bb - m_1 \partial_t \bq$ & 
    $\bb -\frac{1}{T}\mathbf I $ & 
    $\nabla\cdot \bB - m_2 \partial_t \bQ$ &
    $\bB $\\ \hline
Forces &$ \bq $ &
    $\nabla \bq$ &
    $\bQ$ &
    $\nabla \bQ$\\
    \end{tabular}\\
\vskip .21cm
{Table 1. Thermodynamic fluxes and forces}
\end{center}

Let us consider an isotropic material and one dimensional propagation. Then the linear solution of inequality \re{entrpr} is
\begin{eqnarray}
m_1 \partial_t q - \partial_x b &=& -l_1 q, \label{ce1}\\
m_2 \partial_t Q - \partial_x B &=& -k_1 Q + k_{12} \partial_x q, \label{ce2}\\
b- \frac{1}{T} &=& -k_{21} Q + k_{2} \partial_x q, \label{ce3}\\
B &=& n \partial_xQ. \label{ce4},
\end{eqnarray}

where the following conditions are required on the coefficients $l_1, k_1, k_2, k_{12}, n$:
\begin{equation} 
l_1\geq 0, \quad
k_1\geq 0, \quad
k_2\geq 0, \quad 
n\geq 0, \quad  \quad
K=k_1k_2-k_{12}k_{21} \geq 0.
\label{coeffreq}
\end{equation}

Eliminating the current multipliers $b$ and $B$ by substitution, evolution equations for $q$ and $Q$ are obtained. Then we define the relaxation times $\tau_q= \f{m_1}{l_1}$, $\tau_Q= \f{m_2}{k_1}$, the Fourier heat conduction coefficient  $\lambda = \f{1}{T^2 l_1}$, and  material parameters
$\kappa_{21}=\f{k_{21}}{l_1}$, $\kappa_{12}=\f{k_{12}}{k_1}$ . In order to obtain a system that is close to the hyperbolic one of Rational Extended Thermodynamics \cite{DreStr93a}, we simplify the system introducing the condition $n=k_2=0$. The resulting system of equations has been called {\em ballistic-conductive model} \cite{KovVan15a}.

The front side boundary condition is a heat pulse modelled by the following form:
\begin{center}
	$q_0(t) \equiv q(x=0,t)= \left\{ \begin{array}{cc}
	q_{\text{max}}\left(1-\cos\left(2 \pi \cdot \frac{t}{t_p}\right)\right) & 
	\textrm{if } 0<t \leq t_p,\\
	0 & \textrm{if } t>t_p.
	\end{array} \right.  $
\end{center}

At the rear side of the specimen, $x=L$, we consider adiabatic insulation, therefore, $q(x=L,t)=0$. Initially the fields are homogeneous, and the initial conditions are $T(x,t=0)=T_0$ and $q(x,t=0)=0$.

After these preparations we introduce the dimensionless variables  $\hat t, \hat 
x, \hat T, \hat q, \hat Q$ for time, space, temperature, heat 
flux and the internal variables, respectively. 
\begin{eqnarray}
\hat{t} =\frac{\alpha t}{L^2} \quad &\text{with}& \quad
\alpha=\frac{\lambda}{\rho c};  \quad
\hat{x}=\frac{x}{L};\nonumber \\
\hat{T}=\frac{T-T_{0}}{T_{\text{final}}-T_{0}} \quad &\text{with}&\quad
T_{\text{final}}=T_{0}+\frac{\bar{q}_0 t_p}{\rho c L};  \nonumber \\
\hat{q}=\frac{q}{\bar{q}_0} \quad &\text{with}&\quad
\bar{q}_0=\frac{1}{t_p}  \int_{0}^{t_p} q_{0}(t)dt; \nonumber \\
\hat{Q} = \sqrt{\f{\kappa_{12}}{\kappa_{21}}} \bar{q}_0 Q.
\label{ndvar}\end{eqnarray}

The following dimensionless parameters are introduced, corespondingly
\begin{equation}
\hat{\tau}_\Delta =\frac{\alpha t_p}{L^2}; \quad 
\hat{\tau}_q 	  = \frac{\alpha \tau_{q}}{L^2}; \quad   
\hat{\tau}_Q 	  = \frac{\alpha  \tau_{Q}}{L^2}; \quad 
\hat{\kappa} 	  = \f{\sqrt{\kappa_{12} \kappa_{12}}}{L}.
\end{equation}
Finally, we obtain the ballistic-conductive equations in a nondimensional form
\begin{eqnarray}
\hat{\tau}_{\Delta}\partial_{\hat t} \hat T + 
	\partial_{\hat x} \hat q &=& 0 \ ,\nonumber \\
\hat{\tau}_q \partial_{\hat t} \hat q + \hat q + 
    \hat{\tau}_{\Delta}\partial_{\hat x}\hat T + 		
    \hat{\kappa}\partial_{\hat x}\hat Q &=& 0 \ 
    ,\label{nd_balcond}\\
\hat{\tau}_Q \partial_{\hat t}\hat Q +\hat Q + 
	\hat \kappa \partial_{\hat x}\hat q &=& 0 .\nonumber 
\end{eqnarray}
Here, all parameters are nonnegative, according to the inequalities \re{coeffreq}. This set of equations is similar to the set of equations that one can derive from the kinetic theory of phonons \cite{MulRug98b}. 

It is instructive to eliminate $q$ and $Q$ to reach the partial differential equation of temperature:
\begin{equation}
\hat{\tau}_q \hat{\tau}_Q\partial_{\hat t \hat t \hat t} \hat T + 
(\hat{\tau}_q + \hat{\tau}_Q)\partial_{\hat t \hat t} \hat T +  
\partial_{\hat t} \hat T-
(\hat{\tau}_Q + \hat{\kappa}^2)\partial_{\hat t \hat x \hat x} \hat T -
\partial_{\hat x\hat x}\hat T = 0
\end{equation}
One can see that the boundary conditions for this single partial differential equation are not straightforward.

The phase speed is calculated with the help of dispersion relations or hyperbolicity conditions. For instance, the  ballistic propagation speed, $c$, reads
\begin{equation}
c = \sqrt{\frac{\hat{\kappa}^2 + {\hat\tau}_Q}{{\hat\tau}_q {\hat\tau}_Q}}.
\end{equation}
In what follows, we will use this form to adjust the appropriate propagation speed.

\section{Theory versus experiments}

In this section, we reproduce the results of NaF measurements of Jackson and Walker at 13 K (Figure \ref{fig:bal1}, left, middle curve) in the framework of the previously outlined continuum theory of ballistic-conductive system of equations, \re{nd_balcond},  and compare the parameters and modeling results with the performance of the 9-field approximation of the kinetic theory by Dreyer and Struchtrup, and and the complex viscosity calculations of Yanbao Ma \cite{Ma13a}. Our calculations have been performed with the finite difference method described in \cite{KovVan15a}.

\subsection{Results of Dreyer and Struchtrup} The field equations of Extended Rational Thermodynamics are originated in kinetic theory and modelled through phonon-phonon interactions \cite{MulRug98b,DreStr93a, Chen01, Tzou14}. The 9-field approximation results in a set of equations that is similar to ours. The parameters are specific heat, heat conductivity, and only  one of the relaxation times was fitted to the measurements. However, their goal was only a qualitative reproduction of the measurement data:  the ballistic propagation speed could not match because of theoretical reasons. In \re{nd_balcond}, we have one more material parameter, $\hat{\kappa}$, allowing a more accurate modeling of wave phenomena. 

Let us consider Figure 5.1/b from \cite{DreStr93a}. One needs to consider the material parameters of the pure crystal at 13 K to reproduce this simulation. Let the thermal conductivity be $k=13500 \,\frac{W}{mK}$ based on Figure 3/curve B of \cite{JacWal71}, which corresponds to the "less pure" case. Density is $\rho=2866\, \frac{kg}{m^3}$ and specific heat is $c_v=1.8\, \frac{J}{kgK}$ \cite{BarSte05a}, where we considered the temperature dependency from \cite{HarJas71a}. From these, thermal diffusivity can be calculated: $\alpha=\frac{k}{\rho c_v}=2.617\, \frac{m^2}{s}$. In \cite{DreStr93a}, there is no explicit data about the length of the specimen, however, there is a correlation on the top of the figures: $\frac{x}{c \Delta t}=23$, where $x$ is the sample length, $c$ denotes the Debye speed and $\Delta t$ is the pulse length. Here, $\Delta t = 10^{-7}\, s$ is given. $c=\sqrt{\frac{3 \alpha}{\tau_R}}$, one can find  $c=2747.5 \,\frac{m}{s}$, because $\tau_R = 10.4 \Delta t$ is also given. Then one obtains that the length of the sample is $x=6.38 mm$.  The second relaxation time is also given as $\tau = 2.1 \Delta t$. Hence all parameters are reproduced to calculate the dimensionless coefficients of our model, which turn out to be:
\begin{equation*}
\hat \tau_q=0.0686; \quad 
\hat \tau_Q=0.0138; \quad 
\hat \tau_{\Delta}=0.0066; \quad 
\hat \kappa=0.102.
\end{equation*}
  
\begin{figure}
\centering
\includegraphics[width=12cm,height=7cm]{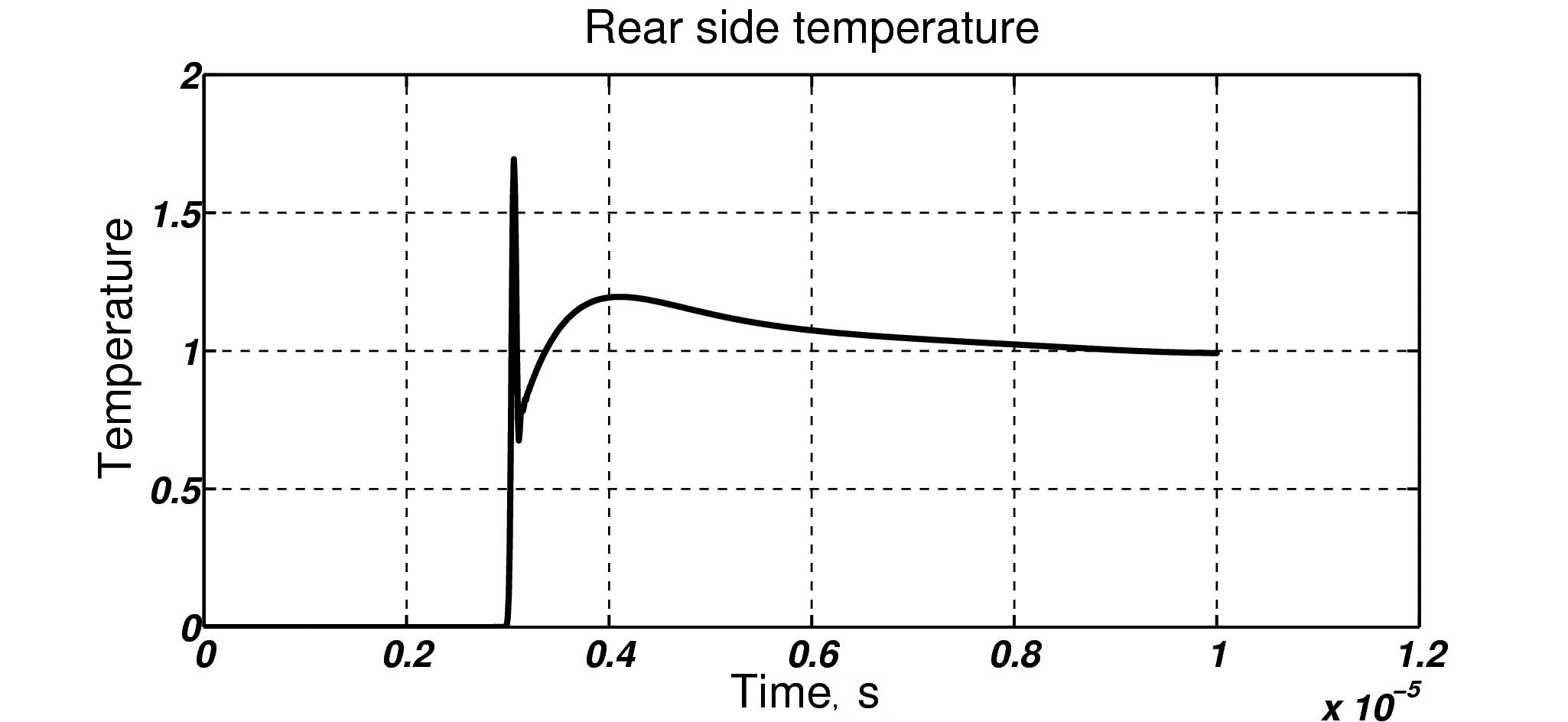}
\caption{NaF experiment: reproduction of the Dreyer--Struchtrup model, Figure 5.1/b in \cite{DreStr93a}. }
\label{fig:bal2}
\end{figure}

Applying adiabatic boundary condition at the rear side of the sample calculation of the time dependency of the rear side temperature results in Figure \ref{fig:bal2}.

\subsubsection{Discussion} Let us note that the length of the crystal does not match with the data from \cite{JacWal71}, where $L=7.9$mm. Moreover, the pulse length, which is not given exactly in the experimental papers, is different from the one  used by Yanbao Ma \cite{Ma13a1}. In their calculations, Dreyer and Struchtrup used Laplace transformations, with environmental temperature at the infinity \cite{DreStr93a}. This assumption is a particular realization of cooling at the rear side of the finite sample. In case of adiabatic conditions, the final temperature depends on the energy content of the pulse and corresponds to $\hat T = 1$.  Therefore, as a summary, we can say that the difference in the material parameters of the two models (the extra $\hat{\kappa}$ coefficient in our case) does not seems to play a role in the qualitative reproduction of their calculation.

\subsection{Results of Yanbao Ma \cite{Ma13a1, Ma13a}}

Yanbao Ma applied  the hydrodynamic theory, where relaxation effects are modeled by complex viscosity \cite{LandauVIeng, Rog71a}. He obtained a good quantitative matching with the experimental results including both ballistic speeds \cite{Ma13a1}. In his case, the duration of the input heat pulse is $0.24 \hbox{\textmu}s$, and the length of the sample is $L=7.9$mm. He does not provide information about the rear side boundary condition. The published relaxation times are $\tau_R=0.937\, \hbox{\textmu}s; \tau=0.248\,\hbox{\textmu}s$ and the speed of sound is also given,  $c=3186 \,\frac{m}{s}$, which is determined from the original measurement data. These papers do not contain information about the other material parameters, thus we use the same values as in the previous case,  $\alpha=2.617\, \frac{m^2}{s}$. Therefore our dimensionless parameters are
\begin{equation*}
\hat \tau_q =0.0393; \quad
\hat \tau_Q=0.0091; \quad
\hat \tau_{\Delta} = 0.01; \quad
\hat \kappa=0.123.
\end{equation*}

\begin{figure}[h]
\centering
\includegraphics[width=12cm,height=7cm]{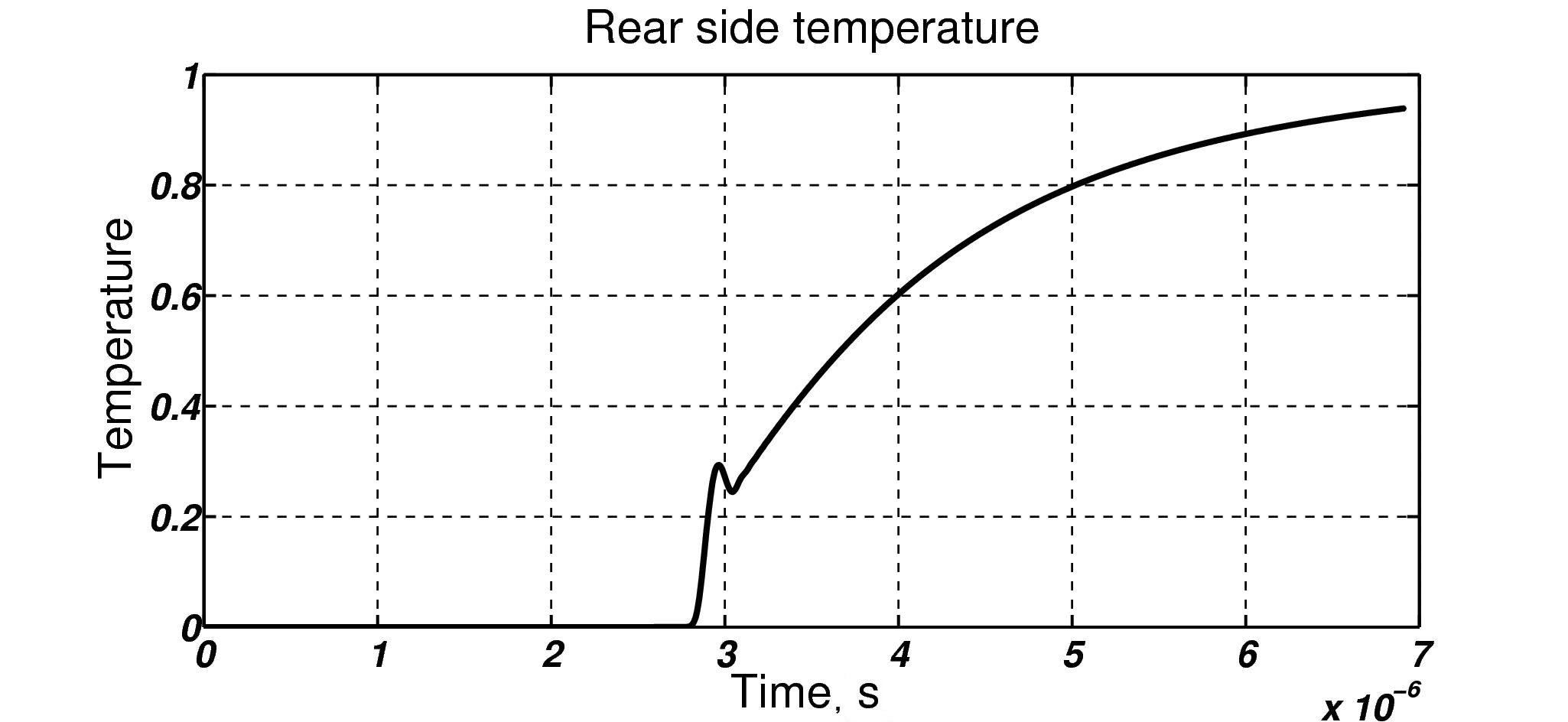}
\caption{NaF experiment: reproduction of the Yanbao Ma model in \cite{Ma13a1}.}
\label{fig:bal3}
\end{figure}

The calculation results in the time dependence of the rear side temperature as shown in Figure \ref{fig:bal3}.

\subsubsection{Discussion} Apparently, these parameters do not reproduce the measurement data; the difference is significant. The structure of the equations is not the same, and the uncertainty in the boundary conditions also seems to be an important factor. 

\subsection{Results of nonequilibrium thermodynamics \cite{KovVan15a}}

\subsubsection{Modification of the model}

According to the inherently two-dimensional arrangement of the measurements with a point-like heat pulse at the front end \cite{McN74t}, it is reasonable to assume that the conservation of the internal energy is violated. Moreover, even if we assume that the thermal insulation at the rear end is not perfect, we cannot obtain sufficient heat loss to explain the diffusive tail of the measurement in \cite{JacWal71a}. Therefore, we introduce a source term, a heat exchange between the bulk crystal and the pulse channel in the balance equation \re{inten_bal} in the following form:
\begin{equation}
\partial_t e + \partial_x  q =-a(T_{ }-T_0),
\label{inten_bal_mod}
\end{equation}
where $a$ is a heat exchange parameter, and $T_0=13$K is the environmental temperature. This is an emulation of the real three-dimensional effect in a one-dimensional framework. 

The dimensionless form of the equation is  
\begin{equation}
{\hat \tau}_\Delta\partial_{\hat t}\hat T + \partial_{\hat x} \hat q +\hat a \hat T =0,
\label{inten_bal_modnd}
\end{equation}
with
\begin{equation}
\hat a =\frac{t_p}{\rho c} a.
\end{equation}

\subsubsection{Material parameters}
Comparing the information from \cite{McN74t} and \cite{JacWal71}, the heat conduction values seem to be contradictory. The crystal in the experiment according to \cite{JacWal71} is the one labelled as "B" in Figure 3 of \cite{JacWal71}, which has the peak thermal conductivity around 15000 $\frac{W}{mK}$. However, McNelly points out regarding Figure 32 in \cite{McN74t} that this crystal is a different one (namely \#7204205W \cite{McN74t}), with peak thermal conductivity value around 12000 $\frac{W}{mK}$. Thus we apply here $\lambda =10200\, \frac{W}{mK}$ instead of the value $13500\,\frac{W}{mK}$ based on the Appendix of thermal conductivity data in \cite{McN74t}. The relaxation times were determined from the transversal ballistic propagation speed and by the proper fit to the second sound peak, and they are $\tau_q=0.355\, \hbox{\textmu}s$ and $ \tau_Q=0.21 \,\hbox{\textmu} s$. Our additional parameters are $\kappa=0.972$mm and the heat exchange coefficient is $a=3.2 \,\frac{W}{mm^3 K} $.

The dimensionless parameters are, accordingly
\begin{equation*}
\hat \tau_q =0.0393;  \quad
\hat \tau_Q=0.0091;  \quad
\hat \tau_{\Delta} = 0.01;  \quad
\hat\kappa=0.123;   \quad
\hat a =0.149.
\end{equation*}

\begin{figure}[h]
\centering
\includegraphics[width=12cm,height=7cm]{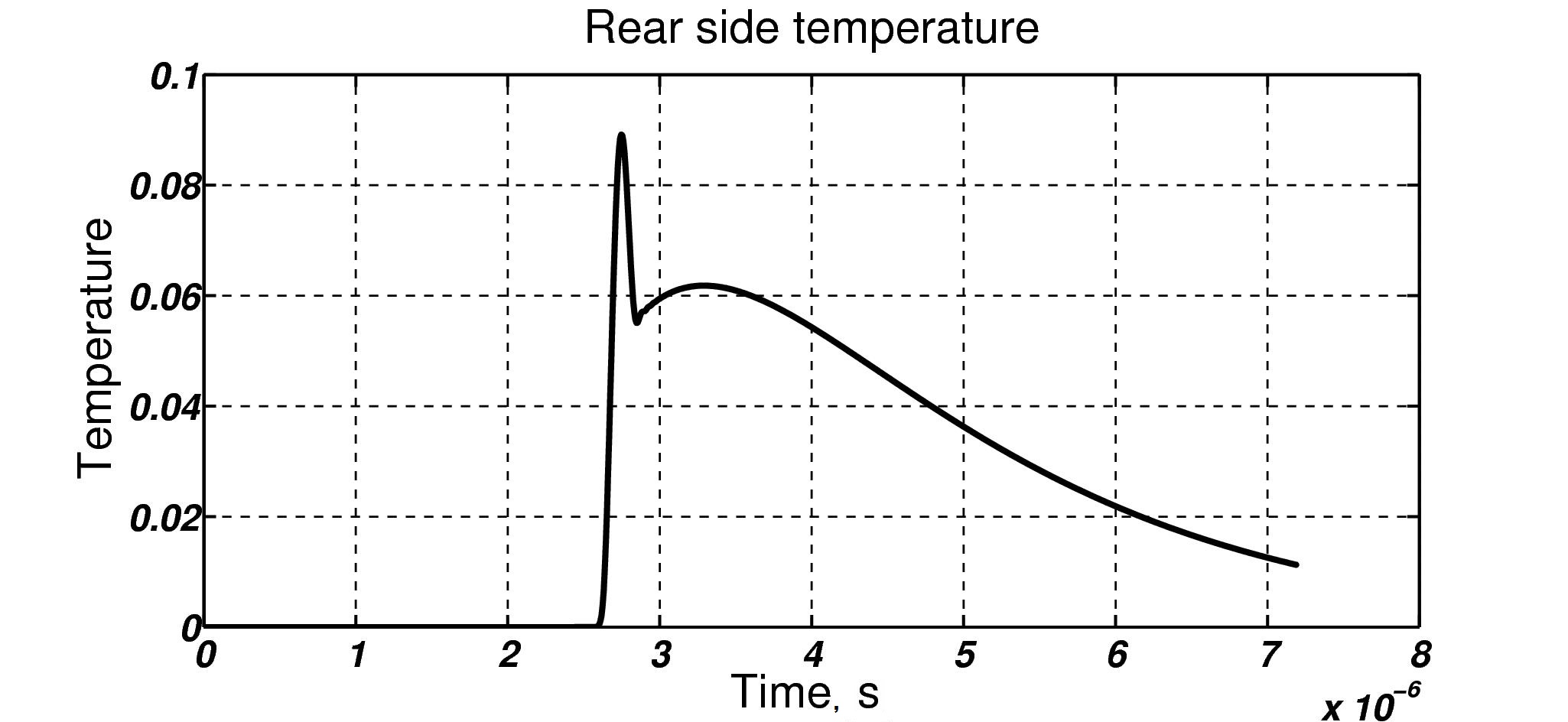}
\caption{NaF experiment: reproduction of the NaF experiment at $13$K in \cite{JacWal71a}. The ratio of amplitudes is calculated without considering the longitudinal wave.}
\label{fig:bal4}
\end{figure}

\begin{figure}[h]
\centering
\includegraphics[width=7cm,height=12cm]{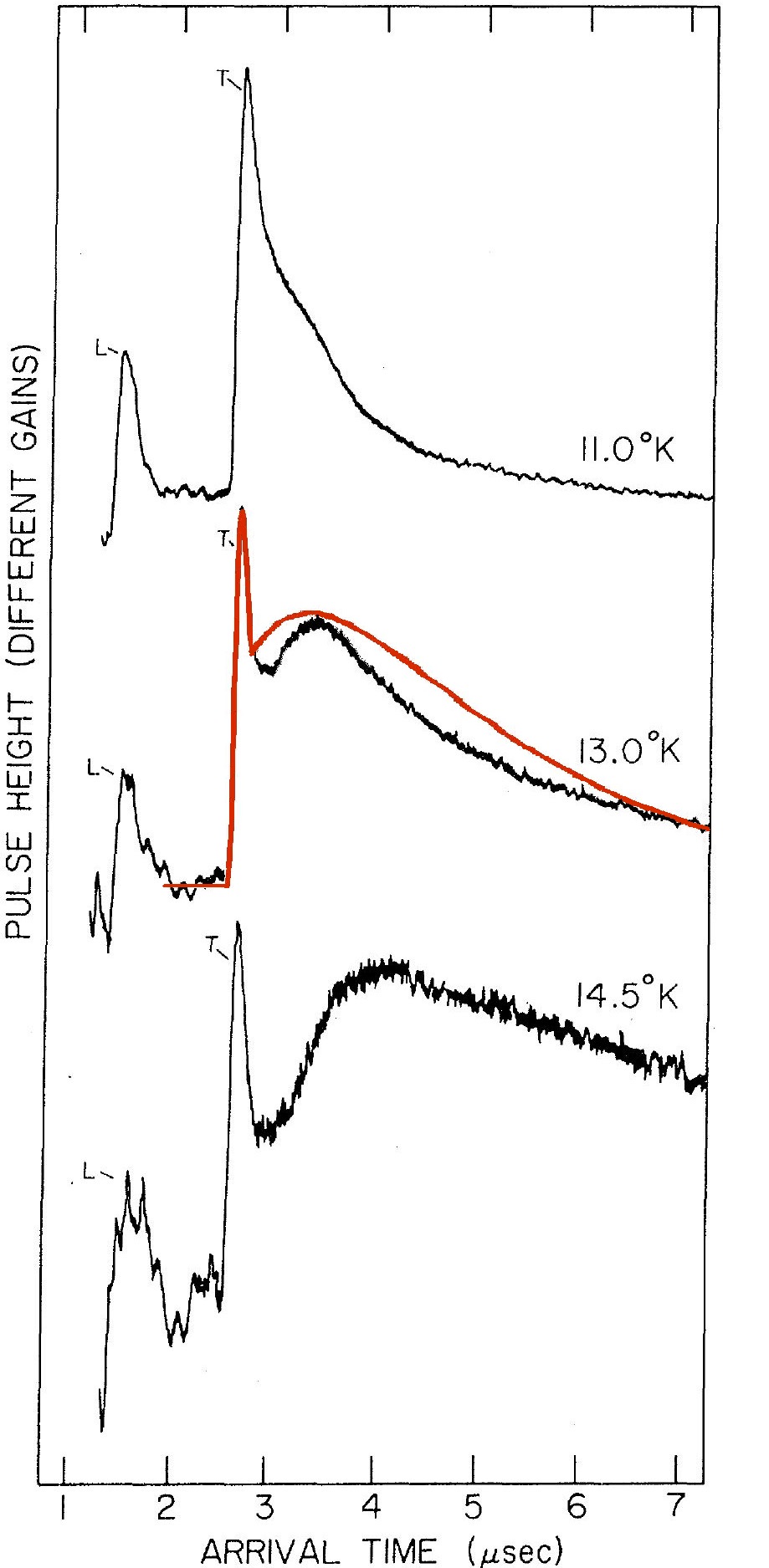}
\caption{NaF experiment: combination of the original figure \cite{JacWal71a} and the simulation (Figure \ref{fig:bal4}) .}
\label{fig:bal5}
\end{figure}

The corresponding solution of the partial differential equation is shown in Figure \ref{fig:bal4}, displayed together with is also combined with the originally measured curve (Figure \ref{fig:bal5}).

Due to the two additional parameters, the slope of the diffusive tail and the ratio of the amplitudes can be calibrated, too. 

\section{Summary and conclusions}

The ballistic-wave-like propagation measurements in NaF by Jackson, Walker and McNelly are considered as the best tests of non-Fourier heat conduction at low temperature. There are several calculations that qualitatively reproduce the time dependence of the rear side temperature. Quantitative reproduction is not an easy task because of the uncertainties of the material parameters and of the boundary conditions. 

In this paper, we have compared an experiment at $13$K and three theories in the framework of our continuum model, published in \cite{KovVan15a}. We found that this theory can be fitted well against the original measurement data, however, it requires different parameters than the theory of Extended Rational Thermodynamics or complex viscosity hydrodynamics suggest. Moreover, the quantitative reproduction of the diffusive tail, considering also realistic boundary conditions at the rear end, requires an additional parameter that characterizes internal cooling. This effect seems to provide a realistic emulation of the real, three-dimensional phenomena with a point-like thermal excitation \cite{McN74t}. However, without the anisotropic heat conduction properties of the crystals, it is impossible to obtain a complete validation. 

\bibliographystyle{unsrt}

\begin{thebibliography}{10}
	
	\bibitem{BotEta16a}
	S. Both, B. Cz\'el, T. F\"ul\"op, Gy. Gr\'of, \'A Gyenis, R. Kov\'acs, P. V\'an, and  J. Verh\'as. 
	\newblock Deviation from the {F}ourier law in room-temperature heat pulse experiments.
	\newblock {\em J Non-Equil Thermody}, 41(1):41-48, 2016.
	
	\bibitem{Tisz38a}
	L.~Tisza.
	\newblock Transport phenomena in {H}elium {II}.
	\newblock {\em Nature}, 141:913, 1938.
	
	\bibitem{Lan41a}
	L. D. Landau.
	\newblock Two-fluid model of liquid {H}elium {II}.
	\newblock {\em J. Phys. USSR}, 5(1):71--90, 1941.
	
	\bibitem{Pes44a}
	V.~Peshkov.
	\newblock Second sound in {H}elium {II}.
	\newblock {\em J. Phys. (Moscow)}, 8:381, 1944.
	
	\bibitem{Cat48a}
	C.~Cattaneo.
	\newblock Sulla conduzione del calore.
	\newblock {\em Atti Sem. Mat. Fis. Univ. Modena}, 3:83--101, 1948.
	
	\bibitem{Ver58a1}
	M.~P. Vernotte.
	\newblock Le paradoxes the la th\'eorie continue e l'\'equation de la chaleur.
	\newblock {\em Comptes rendus hebdomadaires des s\'eances de l'Acad\'emie des
		sciences}, 246:3154--55, 1958.
	
	\bibitem{Max867a}
	J.~C. Maxwell.
	\newblock On the dynamical theory of gases.
	\newblock {\em Philos T R Soc Lond},
	157:49--88, 1867.
	
	\bibitem{GuyKru66a1}
	R.~A. Guyer and J.~A. Krumhansl.
	\newblock Solution of the linearized phonon {B}oltzmann equation.
	\newblock {\em Phys Rev}, 148(2):766--778, 1966.
	
	\bibitem{SelEta16b}
	A. Sellitto, V.A. Cimmelli and D. Jou.
	\newblock {\em Mesoscopic Theories of Heat Transport in Nanosystems}.
	\newblock  Springer International Publishing, 2016.
	
	\bibitem{Zhu16a}
	K. V. Zhukovsky. 
	\newblock Exact solution of {G}uyer–{K}rumhansl type heat equation by operational method.
	\newblock {\em Int J Heat Mass Tran}, 96:132--144, 2016.	
	
	
	\bibitem{DreStr93a}
	W.~Dreyer and H.~Struchtrup.
	\newblock Heat pulse experiments revisited.
	\newblock {\em Continuum Mech Therm}, 5:3--50, 1993.
	
	\bibitem{Rog71a}
	S. J. ~Rogers.
	\newblock Transport of heat and approach to second sound in some isotopically
	pure alkali-halide crystals.
	\newblock {\em Phys Rev B}, 3(4):1440, 1971.
	
	\bibitem{Van15a}
	P.~V\'an.
	\newblock Theories and heat pulse experiments of non-{F}ourier heat conduction.
	\newblock {\em Communications in Applied and Industrial Mathematics}, 2015.
	\newblock arXiv:1501.04234.
	
	\bibitem{McNEta70a}
	T. F. McNelly, S. J. Rogers, D. J. Channin, R. J. Rollefson, W. M. Goubau, G. E.
	Schmidt, J. A. Krumhansl, and R. O. Pohl.
	\newblock Heat pulses in {NaF}: onset of second sound.
	\newblock {\em Phys Rev Lett}, 24(3):100--102, 1970.
	
	\bibitem{BarSte05a}
	S.~Bargmann and P.~Steinmann.
	\newblock Finite element approaches to non-classical heat conduction in solids.
	\newblock {\em Comput. Model. Eng. Sci}, 9(2):133–150,
	2005.
	
	\bibitem{Ma13a1}
	Yanbao Ma.
	\newblock A transient ballistic--diffusive heat conduction model for heat pulse
	propagation in nonmetallic crystals.
	\newblock {\em Int J Heat Mass Tran}, 66:592--602,
	2013.
	
	\bibitem{Nyi91a1}
	B.~Ny\'\i{}ri.
	\newblock On the entropy current.
	\newblock {\em J Non-Equil Thermody}, 16:179--186, 1991.
	
	\bibitem{KovVan15a}
	R.~Kov\'acs and P.~V\'an.
	\newblock Generalized heat conduction in heat pulse experiments.
	\newblock {\em Int J Heat Mass Tran}, 83:613--620,
	2015.
	\newblock arXiv:1409.0313v2.
	
	\bibitem{MulRug98b}
	I.~M\"uller and T.~Ruggeri.
	\newblock {\em Rational Extended Thermodynamics}, volume~37 of {\em Springer
		Tracts in Natural Philosophy}.
	\newblock Springer Verlag, New York-etc., 2nd edition, 1998.
	
	\bibitem{JacWal71a}
	H.~E. Jackson and C.~T. Walker.
	\newblock Thermal conductivity, second sound and phonon-phonon interactions in
	{N}a{F}.
	\newblock {\em Phys Rev B}, 3(4):1428--1439, 1971.
	
	\bibitem{GK66}
	R.~A. Guyer and J.~A. Krumhansl.
	\newblock Thermal conductivity, second sound, and phonon hydrodynamic phenomena
	in nonmetallic crystals.
	\newblock {\em Phys Rev}, 148:778--788, 1966.
	
	\bibitem{JacWalMcN70}
	H.~E. Jackson, C.~T. Walker, and T.~F. McNelly.
	\newblock Second sound in {N}a{F}.
	\newblock {\em Phys Rev Lett}, 25(1):26--28, 1970.
	
	\bibitem{JacWal71}
	H. E. Jackson and C. T. Walker.
	\newblock Thermal conductivity, second sound and phonon-phonon interactions in
	{NaF}.
	\newblock {\em Phys Rev B}, 3(4):1428--1439, 1971.
	
	\bibitem{Ma13a}
	Yanbao Ma.
	\newblock A transient ballistic–diffusive heat conduction model for heat
	pulse propagation in nonmetallic crystals.
	\newblock {\em Int J Heat Mass Tran}, 66:592--602,
	2013.
	
	\bibitem{Chen01}
	G.~Chen.
	\newblock Ballistic-diffusive heat-conduction equations.
	\newblock {\em Phys RevLETT}, 86(11):2297--2300, 2001.
	
	\bibitem{Rog72a}
	S.~J. Rogers.
	\newblock Second sound in solids: the effects of collinear and non-collinear
	three phonon processes.
	\newblock {\em Le Journal de Physique Colloques}, 33(4):4--111, 1972.
	
	\bibitem{Wal63}
	C.~T. Walker.
	\newblock Thermal conductivity of some alkali halides containing {F} centers.
	\newblock {\em Phys Rev}, 132(5):1963--1975, 1963.
	
	\bibitem{McN74t}
	T. ~F. McNelly.
	\newblock Second sound and anharmonic processes in isotopically pure
	alkali-halides. PhD dissertation, \newblock{Cornell University, Ithaca, N.Y},
	\newblock 1974.
	
	\bibitem{Gmelin93}
	E.~Parth{\~a}e and L.~Gmelin.
	\newblock {\em Gmelin Handbook of inorganic and organometallic chemistry:
		TYPIX.. Standardized data and crystal chemical characterization of inorganic
		structure types}, volume~2.
	\newblock Springer-Verlag, 1993.
	
	\bibitem{HarJas71a}
	R.~J. Hardy and S.~S. Jaswal.
	\newblock Velocity of second sound in {N}a{F}.
	\newblock {\em  Phys Rev B}, 3(12):4385--4387, 1971.
	
	\bibitem{ColNew88}
	B.~D. Coleman and D.~C. Newman.
	\newblock Implications of a nonlinearity in the theory of second sound in
	solids.
	\newblock {\em Phys Rev B}, 37(4):1492, 1988.
	
	\bibitem{Verhas97}
	J.~Verhás.
	\newblock {\em Thermodynamics and {R}heology}.
	\newblock Akad\'emiai Kiad\'o-Kluwer Academic Publisher, 1997.
	
	\bibitem{VanFul12}
	P.~Ván and T.~Fülöp.
	\newblock Universality in heat conduction theory -- weakly nonlocal
	thermodynamics.
	\newblock {\em Ann Phys-Berlin}, 524(8):470--478, 2012.
		
	\bibitem{Tzou14}
	D.~Y. Tzou.
	\newblock Longitudinal and transverse phonon transport in dielectric crystals.
	\newblock {\em J Heat Transf}, 136(4):042401, 2014.
	
	\bibitem{LandauVIeng}
	L.~D. Landau and E.~M. Lifshitz.
	\newblock {\em Theoretical Physics. Vol. 6. Hydrodynamics}.
	\newblock Nauka, Moscow, 1986.

    \bibitem{FriCim98m}
    K. ~Frischmuth and V. A. ~Cimmelli.
   \newblock Coupling in thermo-mechanical wave propagation in {NaF} at low temperature.
          \newblock{\em Arch Mech}, 50:703--714, 1998.

	
	
\end{thebibliography}


\end{document}